\documentclass[twocolumn,nofootinbib,prl,floatfix,preprintnumbers,amsmath,amssymb,groupedaddress,superscriptaddress]{revtex4}
\usepackage{dsfont}
\usepackage{graphicx} 
\usepackage{epstopdf} 

\newcommand{\GeV}{\,\mathrm{GeV}}

\newcommand{\pslash}{\!\not\! p}
\newcommand{\figref}[1]{Figure~\ref{#1}}
\newcommand{\definmath}[2] {\def#1{\ifmmode#2\else$#2$\fi}}
\definmath{\invfb}{\mathrm{fb}^{-1}}

\begin{document}   
\title{Measuring the Higgs boson mass in dileptonic $W$-boson decays at hadron colliders}
\date{\today}

\author{Alan J. Barr}
\email{a.barr@physics.ox.ac.uk}
\affiliation{Denys Wilkinson Building, Keble Road, Oxford, OX1 3RH, United Kingdom}

\author{Ben Gripaios}
\email{gripaios@cern.ch}  
\affiliation{CERN PH-TH, Geneva 23, 1211 Switzerland}

\author{Christopher G. Lester}
\email{lester@hep.phy.cam.ac.uk}
\affiliation{Cavendish Laboratory, Dept of Physics, JJ Thomson Avenue,
Cambridge, CB3 0HE, United Kingdom}

\preprint{Cavendish-HEP-09/04}
\begin{abstract} 
It is expected that hadron collider measurements of the Higgs boson mass using the decay
$h\rightarrow W^+W^-,$ followed by the leptonic decay of each
$W$-boson, will be performed by fitting the shape of a
distribution that is sensitive to the Higgs mass.  We demonstrate
that the variable most commonly used to measure the Higgs mass in this
channel is not optimal as it contains an unnecessary and even
counter-productive approximation.  We remove that approximation,
without introducing any cost in complexity, and demonstrate that the
new variable is a clear improvement over the old: its performance is
never worse, and in some cases (particularly the high Higgs mass
region) it might reduce the fit uncertainty on the Higgs mass 
in that channel by a factor approaching two.

\end{abstract}   
\maketitle 
\subsection{Introduction}
The mass of the Higgs boson is the last unknown parameter of the Standard Model. 
Here, we present a method
to measure it at a hadron collider, assuming the Higgs exists and is sufficiently massive ($m_h \gtrsim 130 \GeV$) that it decays predominantly to $W$-bosons.
The method is based on the transverse mass observable, $m_T$, that was originally used to measure the masses of the $W$-bosons themselves, via their decays, $W\rightarrow l\nu$, to a lepton and a neutrino. There, since the neutrino is invisible in a detector, one cannot simply reconstruct the mass of the parent $W$ from the invariant mass of the $l\nu$ daughter system; the transverse mass $m_T$ circumvents this problem. Similarly, in the case of Higgs decays to two $W$s (one or more of which may be significantly off-mass-shell), 
then if the $W$s subsequently decay leptonically to $l\nu$, we end up with two invisible neutrinos in the final state. We will describe a generalization of $m_T$ whose distribution features an edge, which will enable us to extract $m_h$ directly. We believe that the method both complements, and improves upon, existing strategies \cite{Aad:2009wy,cmsphystdr,cmshww} for measuring $m_h$ in this channel, and we encourage experiments to make use of it. The distribution should also aid ongoing Higgs searches at the Tevatron \cite{cdfhww,d0hww}. 

We also briefly discuss potential applications to mass measurement of other particles at the LHC, for example new resonances (such as Kaluza-Klein gluons from an extra dimension) that decay to $t\overline{t}$, as well as the lightest stable superpartner (LSP) in supersymmetric theories.

The original application of the transverse mass was in measurement of $m_W$\cite{Arnison:1983rp,Arnison:1983zy,Banner:1983jy}. We define
\begin{gather}\label{mT}
m_T^2 \equiv m_v^2 + m_i^2 + 2 (e_v e_i - \mathbf{p}_v \cdot \mathbf{p}_i),
\end{gather}
where $\mathbf{p}$ is the momentum transverse to the beam, $e=\sqrt{\mathbf{p}\cdot \mathbf{p} + m^2}$ denotes the transverse energy, and $v$ and $i$ label the visible and invisible decay products respectively,
(a charged lepton and a neutrino in the case at hand).

This definition of $m_T$ has two desirable features: first, since the mass of the neutrino is unknown, but negligible, and the transverse momentum of the neutrino
can be inferred from the missing transverse momentum in the event, $m_T$ is indeed an observable; second, $m_T$ is always bounded above by the mass $m_W$ of the parent $W$. This is easily shown using the invariant mass constraint
\begin{gather}
m_W^2 = m_v^2 + m_i^2 + 2 (E_v E_i - \mathbf{p}_v \cdot \mathbf{p}_i - q_v q_i),
\end{gather}
where $q$ is the longitudinal momentum and $E = \sqrt{q^2 + \mathbf{p}\cdot \mathbf{p} + m^2}$ is the energy, together with the lemma
\begin{gather}
E_v E_i - q_v q_i \geq e_i e_v,
\end{gather}
with equality at $E_v q_i = E_i q_v$, which the reader may easily prove for himself. Thus, by computing the distribution of $m_T$ in many events, 
$m_W$ appears as the upper endpoint. (In practice, the finite decay width of the $W$ and other effects lead to $m_W$ appearing as a Jacobian peak in the data.)

Recently, a number of generalizations of $m_T$ have appeared \cite{Lester:1999tx,Barr:2003rg,Lester:2007fq,Cho:2007qv,Gripaios:2007is,Barr:2007hy,Cho:2007dh,Tovey:2008ui,Serna:2008zk,Burns:2008va}, with diverse applications for LHC mass measurements.
They include generalizations to: decays with multiple visible daughters; decays with a massive invisible daughter (such as a DM candidate); and decays of pair-produced parent particles. The last of these has already been used to measure the mass of the top quark in the process $t\overline{t} \rightarrow b\overline{b}W^+ W^-\rightarrow b\overline{b} l^+ l^- \nu \overline{\nu}$ at the Tevatron \cite{cdfmt2}.
\subsection{More Invisibles}
There is one other generalization that can be made, which is to situations where a single 
decay in itself contains more than one invisible daughter. Practical examples include the single Higgs decay $h \rightarrow WW^{(*)}\rightarrow \ell^+\ell^- \nu\bar\nu$, the decay of new resonances (such as a Kaluza-Klein gluon) to $t \overline{t}$, followed by a semi-leptonic decay of each top, $t \rightarrow bW \rightarrow bl\nu$, or pair decays in supersymmetric
theories with both the lightest superpartner and neutrinos in the final state. 
To generalize $m_T$ to such a situation, the obvious thing to do is to replace $m_i$ in (\ref{mT}) by the invariant mass of the invisible system \cite{Barr:2002ex}.
Now, in any event, $m_i$ goes unobserved, but it is useful nevertheless to consider its properties. A first observation is that $m_i$, though a relativistic invariant, now varies from event to event, taking values on some real, positive interval. The endpoints of this interval, $m_{i\lessgtr}$, are fixed by the particular decay topology. 
For example, if a parent of mass $m_0$ undergoes a pointlike three-body decay to one visible particle of mass $m_v$ and two massless invisible particles, the lower and upper endpoints are given by $m_{i<} = 0$ and 
$m_{i>} = m_{0}- m_{v}$, respectively, whereas if the decay involves an intermediate resonance of mass $m_I$, they are given by $m_{i<} = 0 $ and $m_{i>} = \sqrt{(m_0^2-m_I^2)(m_I^2-m_v^2)/m_I^2}$.

What is more, it is easy to show that $m_T$ is a monotonically increasing function of $m_i^2$. We thus have the chain of inequalities
\begin{gather} \label{chain}
m_T (m_i = m_{i<}) \leq m_T (m_i) \leq m_0.
\end{gather}
If $m_{i<}$ is known, then $m_T (m_{i<})$ is an observable that is bounded above by $m_0$; if $m_{i<}$ is unknown, we can determine it using a generalization of the kink method described in \cite{Cho:2007qv,Gripaios:2007is,Barr:2007hy,Cho:2007dh}.
\subsubsection{Higgs Decays}
For the Higgs decay $h \rightarrow WW^{(*)}\rightarrow \ell^+\ell^- \nu\bar\nu$, it is simple enough to show that $m_{i<} = 0$, 
when we ignore the mass of the neutrinos.
To wit, consider the on-shell decay $h \rightarrow WW\rightarrow \ell^+\ell^- \nu\bar\nu$, with $h$ at rest in the laboratory, in which the two $W$s are emitted back-to-back. Then let the two $W$s decay such that the neutrinos are emitted parallel to each other (not anti-parallel). In this configuration, $m_i = 0$. Since $m_i$ is positive semi-definite, $m_{i<} = 0$. Similar arguments apply to the off-shell decay $h \rightarrow WW^*$.

That the inequalities in (\ref{chain}) can be made into equalities also follows from the existence of these kinematic configurations. 
Thus, by computing
\begin{gather}\label{eq:mt_true}
(m_T^\mathrm{true})^2 \equiv m_T^2 (m_i = 0) = m_v^2 + 2 (e_v |\mathbf{p}_i| - \mathbf{p}_v \cdot \mathbf{p}_i),
\end{gather}
in many events, we should obtain a distribution in $m_T^\mathrm{true}$ whose endpoint yields the mass of the Higgs boson. 
Since the observable defined in \eqref{eq:mt_true} is truly bounded above by $m_0$, we  distinguish it from other transverse-mass-like observables by giving it the label  $m_T^\mathrm{true}$.

In work to date \cite{Rainwater:1999sd,Asai:2004ws,Aad:2009wy}, an alternative transverse mass has been used,
\begin{gather}\label{eq:mt_atlas}
m_T^\mathrm{approx} \equiv m_T(m_i = m_v) .
\end{gather}
The justification for replacing the unknown $m_i$ by the observable $m_v$ in those papers is that
for Higgs bosons with masses close to $2 m_W$ and produced at or near threshold,
each $W$ boson will decay almost at rest, therefore $m_i \approx m_v$. We note though that $m_T^\mathrm{approx}$ is not bounded above by $m_0$.
Not knowing $m_i$ and without using the above approximation, the best lower limit we can place on $m_h$ will be with
the true transverse mass \eqref{eq:mt_true}.

\subsubsection{$h \rightarrow WW^{(*)}$ simulation}

To investigate the relative performance of the alternative transverse mass variables
\eqref{eq:mt_true} and \eqref{eq:mt_atlas}
we use the {\tt HERWIG} 6.505 \cite{Corcella:2002jc,Marchesini:1991ch} Monte
Carlo generator, with LHC beam conditions ($\sqrt{s}=14~\rm{TeV}$).
Our version of the generator includes the fix to the $h \to WW^{(*)}$ 
spin correlations described in \cite{herwig-bug-fix}.
Our simulations do not include all corrections from higher orders in 
$\alpha_s$ (see e.g. \cite{Anastasiou:2008ik} for a comparison). 
These will be important to consider when later comparing against real experimental distributions.

We generate unweighted events for Standard Model Higgs boson production ($gg\rightarrow h$) 
and for the dominant background, $q \bar{q} \rightarrow WW$.\footnote{Other backgrounds, such as $Z\rightarrow 2\tau$, are rendered sub-dominant by the cuts discussed below \cite{Aad:2009wy}.}
Final state hadrons with $p_T > 0.5$~GeV and pseudorapidity $|\eta|<5$  
are clustered into jets using the longitudinally invariant $k_T$ clustering algorithm 
for hadron-hadron collisions \cite{Catani:1993hr} in the inclusive mode \cite{Ellis:1993tq} with $R=1.0$.
The missing transverse momentum ${\bf p}_i$ is calculated from the vector sum of the transverse momenta of 
the neutrinos. No detector simulation is applied in this paper. 
Detector effects should provide a relatively small correction since lepton momenta are very well measured at these energies 
\cite{Aad:2009wy,cmsphystdr}, and the dominant contribution to 
the missing transverse momentum ${\bf p}_i$ will be from recoil against well-measured leptons.
There will be some additional smearing of ${\bf p}_i$ from mismeasured and out-of-acceptance hadrons but
such corrections are small when the hadronic transverse energy in the event is small \cite{Aad:2009wy}.

Selection cuts are applied based on \cite{Aad:2009wy}, requiring:
\begin{itemize}
\item{Exactly two leptons $\ell\in\{e,\mu\}$ with $p_T > 15$~GeV and $|\eta|<2.5$}
\item{Missing transverse momentum, $\pslash_T > 30$~GeV}
\item{ $12~\GeV < m_{\ell\ell} < 300~\GeV$}
\item{No jet with $p_T > 20$~GeV }
\item{$Z\rightarrow \tau\tau$ rejection: the event was rejected if
  $|m_{\tau\tau}-m_{Z}|<25~\GeV$ and 
  $0<x_i<1$ for both $i\in \{1,2\}$\footnote{The variable $x_i$ is the 
  momentum fraction of the $i$th tau carried by its daughter lepton and $m_{\tau\tau}$ is the di-tau invariant mass.
  They are calculated using the approximation that each $\tau$ was collinear with its daughter lepton.}}
\item{Relative azimuth $\Delta\phi_{\ell\ell} < \Delta\phi_{\ell\ell}^{\rm max} $}
\item{Transverse momentum of the $W$ pair system, $p_{T\,WW} > p_{T\,WW}^{\rm min}$}
\end{itemize}
As has been done in previous studies \cite{Aad:2009wy}, we optimize the values of the latter two cuts, 
$\Delta\phi_{\ell\ell}^{\rm max}$ and $p_{T\,WW}^{\rm min}$
for each Higgs boson mass.
In this case we select the values which would be predicted to best constrain $m_h$ --
experimentally one would select cuts which would give the best expected measurement once an approximate Higgs boson mass was known. 
The optimal values ranged from 1.4 to 2.4 for $\Delta\phi_{\ell\ell}^{\rm max}$ and 0 to 10~GeV for $p_{T\,WW}^{\rm min}$. 

\begin{figure}[t]
 \begin{center}
  \includegraphics[width=0.9\linewidth,height=0.45\linewidth]{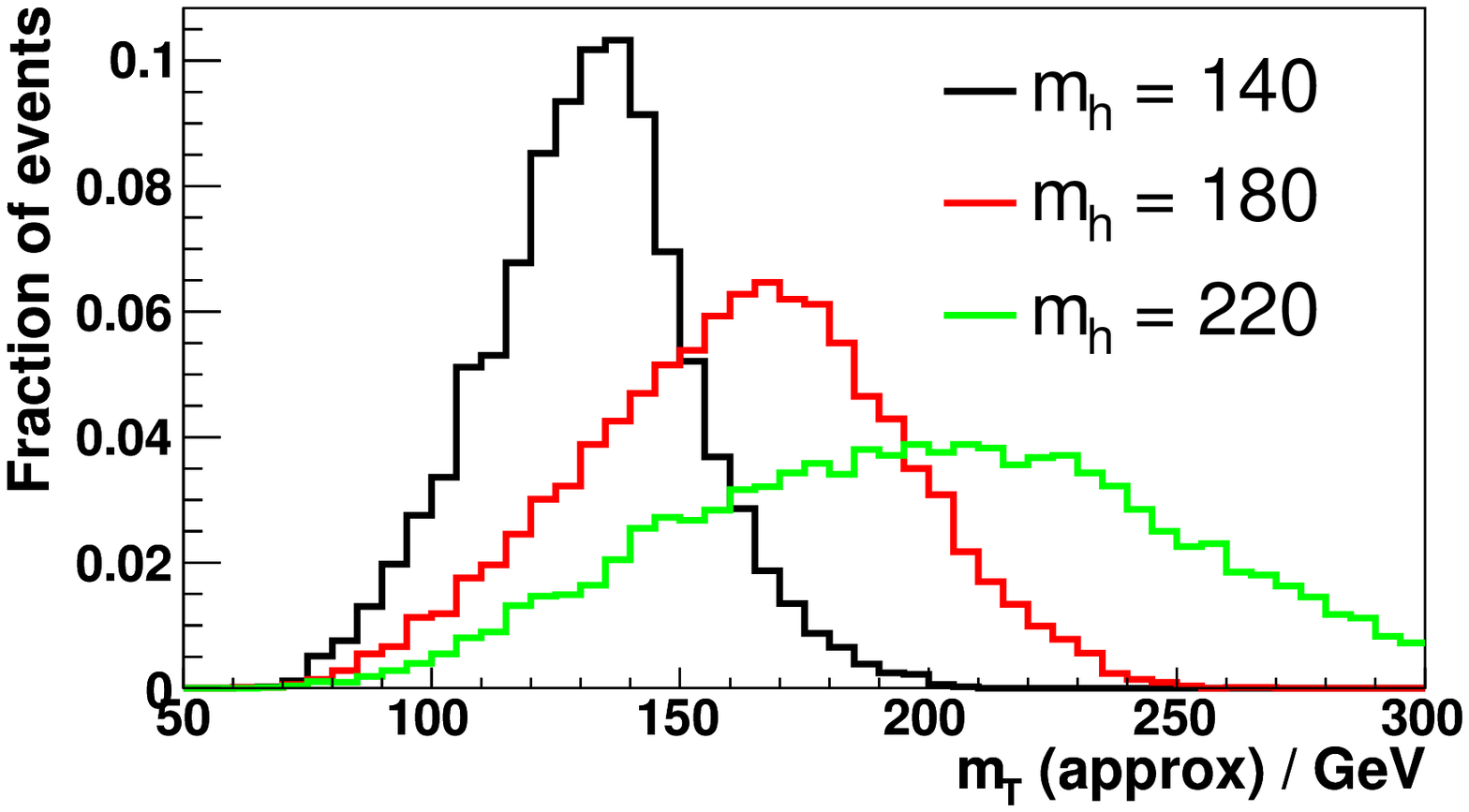}
  \includegraphics[width=0.9\linewidth,height=0.45\linewidth]{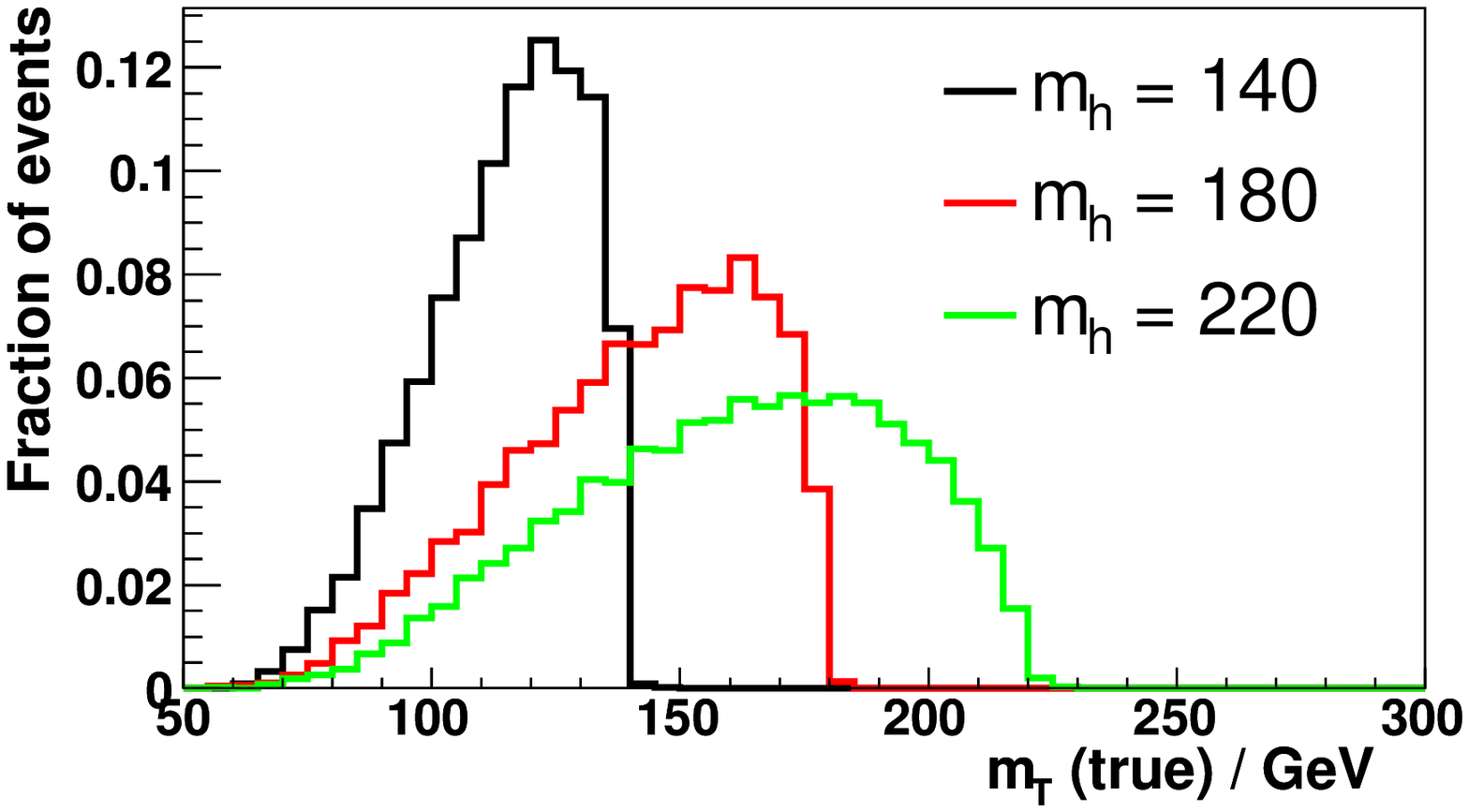}
\caption{
Signal-only distributions of $m_T^\mathrm{approx}$ (top) and $m_T^\mathrm{true}$ (bottom)
for various values of $m_h$ (in GeV). 
No cuts on $\Delta\phi_{\ell\ell}^{\rm max}$ and $p_{T\,WW}^{\rm min}$ have been applied.
}
\label{fig:example_distributions}
\end{center}
\end{figure}

\begin{figure}
 \begin{center}
  \includegraphics[width=1.1\linewidth,height=0.65\linewidth]{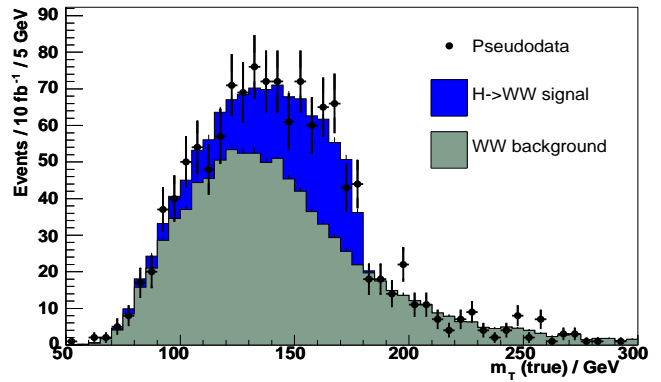}
\caption{
Example pseudo-experiment $m_T^\mathrm{true}$ distribution (points with error bars) 
and model distribution (shaded histograms) for integrated luminosity of 10~\invfb.
The plot includes the $WW$ background and is made for $m_h = m_h^\mathrm{model} = 180~\GeV$, 
$\Delta\phi_{\ell\ell}^{\rm max} = 1.8 $ and $p_{T\,WW}^{\rm min} = 10~\GeV$.
}
\label{fig:T_example}
\end{center}
\end{figure}

\begin{figure}
 \begin{center}
  \includegraphics[width=1.1\linewidth]{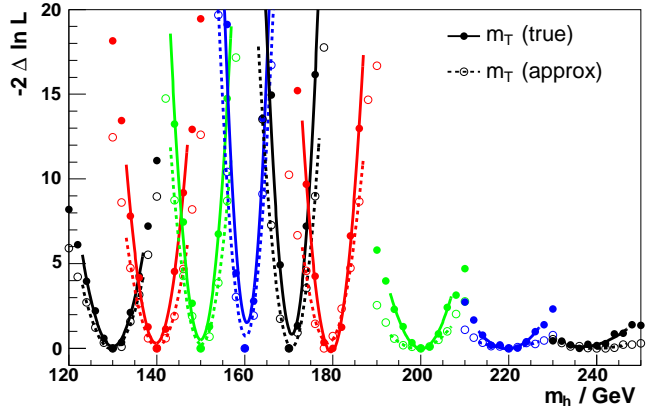}
\caption{
Relative log likelihood distributions for various Higgs boson masses for each of several different input masses
and for both $m_T^{\mathrm{approx}}$ (dashed) and $m_T^{\mathrm{true}}$ (solid). 
The points correspond to integrated luminosity of 10~\invfb, and are plotted for 
$m_h \in \{$130, 140, 150, 160, 170, 180, 200, 220, 240$\}~\GeV$.
}
\label{fig:lnlik_opt}
\end{center}
\end{figure}

\begin{figure}
 \begin{center}
  \includegraphics[width=1.\linewidth,height=0.8\linewidth]{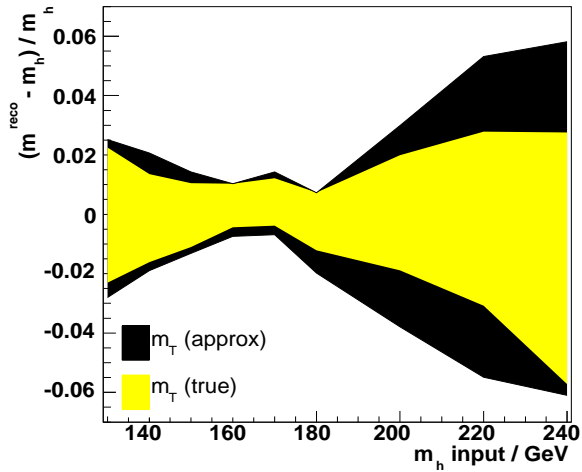}
\caption{
The bands show the fractional uncertainty with which one could expect to measure $m_h$ with  
$m_T^{\mathrm{approx}}$ (black) and $m_T^{\mathrm{true}}$ (shaded)
as a function of $m_h$.
The integrated luminosity simulated is 10~\invfb.
}
\label{fig:lnlik_opt_frac}
\end{center}
\end{figure}

Both observables correlate with $m_h$ (see \figref{fig:example_distributions}),
so it is possible to make a mass measurement with either. 
However $m_T^\mathrm{approx}$ does not provide a strict event-by-event lower bound on $m_h$,
whereas the kinematic endpoint of the $m_T^\mathrm{true}$ distribution shows a clear edge at $m_h$. 

To examine the relative performance of the two variables, 
we generate distributions of them for various choices of $m_h$.
This is done for twenty independent pseudo-experiments (including both signal and the dominant $WW$ background contributions),
each corresponding to integrated luminosity of 10~\invfb.
Each pseudo-experiment is compared to (signal and $WW$ background) 
model distributions with differing hypotheses of 
$m_h^{\rm model}$.

An example pseudo-experiment distribution for $m_T^\mathrm{true}$ is shown in \figref{fig:T_example}.
For each pseudo-experiment the binned log likelihood of the data is calculated.
Each likelihood is maximised over the normalisations of the model $h\to WW$ signal and $WW$ background distributions, 
reflecting our uncertainty in the cross-sections and luminosity\footnote{While uncertainties in the {\em shapes} of these distributions 
are also likely to be important, they are difficult to estimate without collision data and so are not considered in this paper. 
We note that shape effects are likely to be more detrimental for $m_T^{\mathrm{approx}}$; the position of the kinematic edge in $m_T^{\mathrm{true}}$
should be robust against uncertainties in smoothly-varying background parameters.}:
\begin{gather*}
\log\mathcal{L}(m_h, m_h^\mathrm{model}) =
     \left< 
        \max_{\substack{f_\mathrm{SIG} \\ f_\mathrm{BG}}}
              \sum_i \log \mathcal{L}^P\left( n^\mathrm{trial}_i; x_i \right)
     \right>_\mathrm{trials}
\end{gather*} 
where the sum is over histogram bins, $\mathcal{L}^P(n;x)$ is the Poisson likelihood and 
$x_i(f_\mathrm{SIG}, f_\mathrm{BG}, m_h^{\rm model})$
is the expected number of events if signal and background cross sections are $f$ times their leading-order Monte Carlo predictions.
The angle brackets indicate an average over the twenty pseudo-experiments.

The resulting curves of $-2\Delta\log\mathcal{L}$ are plotted in \figref{fig:lnlik_opt},
where $\Delta$ indicates the difference from the minimum value.
The relative precision with which each method can be expected to measure the Higgs boson mass is 
determined from a quadratic fit to $-2\Delta\log\mathcal{L}$ around the minimum.
The fractional uncertainties (\figref{fig:lnlik_opt_frac})
show that the true transverse mass performs somewhat better than the approximate version for all $m_h$, 
so there appears to be no advantage in making the approximation $m_i\approx m_v$.
When $m_h > 2m_W$ there is a significant penalty to pay for assuming $m_i\approx m_v$
--- the true transverse mass provides the higher-precision measurement.

The absolute uncertainties (for both variables) will obviously be somewhat broadened when
experimental resolution and sub-leading backgrounds are included.
While such detailed simulations are beyond the scope of this paper, 
we project that the desirable properties of $m_T^\mathrm{true}$ will mean it
is also the more appropriate variable in the real world.

One might also expect $m_T^\mathrm{true}$ to be a good selection
variable for Higgs boson discovery and for measuring the product of cross-section and branching ratio for Higgs production and di-leptonic decay, by counting the number of signal events. Indeed, as discussed in the Appendix, we find that $m_T^\mathrm{true}$ again gives an improvement, albeit a slight one, over $m_T^\mathrm{approx}$ in both cases.
\subsubsection{Other Applications}
There are many other possible decay processes at the LHC involving multiple invisible daughters, to which similar methods might be applied. One is to decays of
new resonances, such as a Kaluza-Klein gluon from an extra dimension \cite{Agashe:2003zs}, in the $t\overline{t}$ channel, followed by semi-leptonic decays of the tops. For heavy resonances (existing constraints suggest that a KK gluon should be multi-TeV, for example), the approximation will certainly be inappropriate.

A second example is supersymmetric decays involving both the LSP and neutrinos. There, we do not know the
mass of the LSP and we are forced to resort to a kink-based
method, as in \cite{Cho:2007qv,Gripaios:2007is,Barr:2007hy,Cho:2007dh}. 
\subsection{Conclusions}
There seems to be no advantage in using the approximate version of the transverse mass -- 
whether for Higgs boson discovery, for mass determination or for measuring event rates.
Indeed our simulations show that the approximation is often counter-productive, particularly if the objective is 
to make a Higgs boson mass measurement and especially when $m_h>2 m_W$.
The true transverse mass is easy to calculate, and (unlike the approximate version) 
provides an event-by-event lower bound on $m_h$.

These results should be cross-checked with more detailed studies with: 
full detector simulation; more sophisticated models for the signal and background distribution shape uncertainties;
and with calculations to higher orders in $\alpha_s$.\footnote{A subsequent study with full detector simulation \cite{private} confirms our results and, in addition, suggests that the
true transverse mass appears to have the advantage over $m_T^\mathrm{approx}$ of being less correlated
with $\Delta\phi_{\ell\ell}$.}
Future work should also consider the case of Higgs boson production via vector boson fusion
for which one might expect rather similar results.

Other examples of processes where this generalization of $m_T$ could be used include
Kaluza-Klein gluon decays $g_{KK} \rightarrow t\bar{t}$, where the top quarks decay via leptonic $W$ bosons,  
and supersymmetric decays involving neutrinos, such as $\tilde{\chi}^+_1\rightarrow \ell \nu \tilde{\chi}_1^0$.

Although we have focussed our attention here on the decay $h\rightarrow WW$, it is worth remarking that, in the case of an Standard Model Higgs boson with $m_h > 2 m_Z$,
the decay channel $h \rightarrow 2Z \rightarrow 4l$ will allow the Higgs mass to be measured at the {\em per mille} level. Nevertheless, the decay $h\rightarrow WW$
would provide an important corroborative measurement.

\begin{acknowledgments}
We are grateful to Bryan Webber for making us aware of
\cite{herwig-bug-fix}, to other members of the Cambridge Supersymmetry
Working Group, and also to Alessandro Nisati, Bruce Mellado, Bill Quayle and a referee for additional helpful comments.
AJB thanks Chris Hays and members of the Oxford Dalitz Institute for useful discussions.
This work was supported by the Science and Technology Research Council of the United Kingdom.
\end{acknowledgments}
\appendix
\subsection{Appendix: Higgs discovery and branching ratio}
\begin{figure}
 \begin{center}
  \includegraphics[width=0.9\linewidth,height=0.65\linewidth]{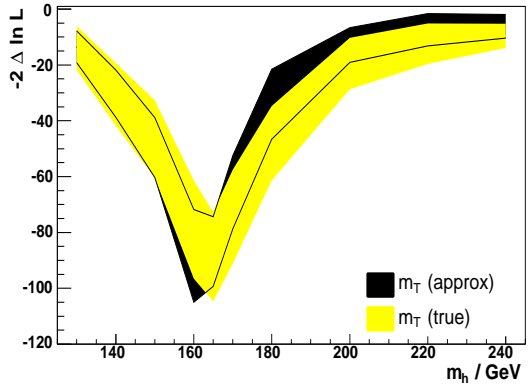}
\caption{
Higgs boson discovery potential as a function of $m_h$.
The center of each band indicates the difference in log likelihood 
between models with and without a Higgs Boson contribution.
Lower values correspond to better discovery potential.
The half-width of the each band gives the root-mean-squared over trial samples.
The integrated luminosity simulated is 10~\invfb.
}
\label{fig:lnlik_discovery}
\end{center}
\end{figure}

\begin{figure}
 \begin{center}
  \includegraphics[width=0.9\linewidth,height=0.65\linewidth]{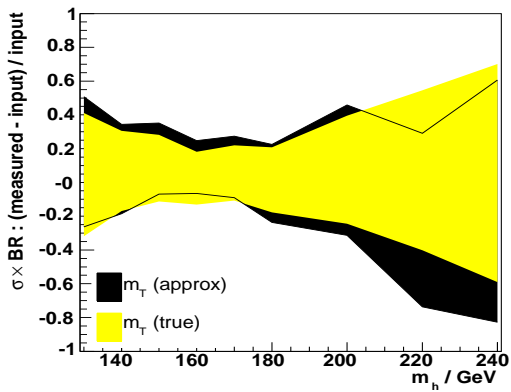}
\caption{
Precision with which one could expect to measure the Higgs boson 
cross-section times branching ratio to $WW^{(*)}$ as a function of $m_h$.
The integrated luminosity simulated is 10~\invfb.
}
\label{fig:width}
\end{center}
\end{figure}

To quantify the Higgs boson discovery potential using $m_T^\mathrm{true}$ as a selection variable, we calculate the log likelihood difference
\begin{equation*}
   -2\Delta\log\mathcal{L} = 
       -2 \left<
           \log \mathcal{L}
               - \log \mathcal{L}^{\not h} 
       \right>_\mathrm{trials} ,
\end{equation*}
where $\mathcal{L}^{\not h}$ is the likelihood of the trial data when the model contains no Higgs boson contribution.
$\mathcal{L}$ is maximised over $f_\mathrm{SIG}$, $ f_\mathrm{BG}$ and $m_h^\mathrm{model}$; 
$\mathcal{L}^{\not h}$ is maximised over $f_\mathrm{BG}$.

Plots of this $-2\Delta\log\mathcal{L}$ are shown as a function of $m_h$ in \figref{fig:lnlik_discovery}.
The absolute numbers are optimistic, since the discovery potential will be reduced 
by subdominant backgrounds and detector resolution,
but the relative performance of the two variables is meaningful. 
One can see that the plots are rather similar, but there may be a
small advantage in using $m_T^{\mathrm{true}}$ rather than 
$m_T^{\mathrm{approx}}$ when $m_h > 2 m_W$.

The branching ratio of the Higgs boson to $W$ boson pairs is another parameter of significant interest.
The relative precision with which one could measure the number of signal events,
which is proportional to $\sigma(pp\rightarrow h)\times \mathrm{BR}(h\rightarrow WW^{(*)})$,
was determined from quadratic fits to 
\begin{gather*}
\log\mathcal{L}(m_h,\ f_\mathrm{SIG}) =
     \left< 
        \max_{f_\mathrm{BG},\ m_h^\mathrm{model}}
              \sum_i \log \mathcal{L}^P\left( n^\mathrm{trial}_i; x_i \right) 
     \right>_\mathrm{trials} 
\end{gather*}
for various $m_h$.
The resulting fractional uncertainty bands are shown in \figref{fig:width}.
There is again an advantage in using $m_T^\mathrm{true}$,
though in this case it is slight.
\bibliography{hww1}
\end{document}